\documentclass{article}
\usepackage{amsmath,amsbsy,amsfonts}

\begin{document}
\newcommand{\nn}{\nonumber \\}
\newcommand{\eqqref}[1]{ Eq. (\ref{#1})}
\newcommand{\eps}{\ensuremath{\varepsilon}}
\newcommand{\me}{\mathrm{e}}
\newcommand{\im}{\ensuremath{\mathrm{i}}}
\newcommand{\bra}[1]{\langle #1 |}
\newcommand{\ket}[1]{| #1 \rangle}
\newcommand{\Bra}[1]{\left\langle #1 \left|}
\newcommand{\Ket}[1]{\right| #1 \right\rangle}
\newcommand{\bigbra}[1] {\big\langle #1\big|}
\newcommand{\bigket}[1] {\big|#1\big\rangle}
\newcommand{\Bigbra}[1] {\Big\langle #1\Big|}
\newcommand{\Bigket}[1] {\Big|#1\Big\rangle}
\newcommand{\dif}{\ensuremath{\mathrm{d}}}
\newcommand{\difbx}{\dif^3\bx}
\newcommand{\difbk}{\dif^3\bk}
\newcommand{\difbp}{\dif^3\bp}
\newcommand{\intD}[1]{\int\frac{\dif^3 #1}{(2\pi)^3}}
\newcommand{\intDD}[1]{\int\frac{\dif^4 #1}{(2\pi)^4}}
\newcommand{\intdinf}[1]{\int_{-\infty}^\infty\frac{\dif #1}{2\pi}}
\newcommand{\intbx}{\int\dif^3\bx}
\newcommand{\bx}{\boldsymbol{x}}
\newcommand{\br}{\boldsymbol{r}}
\newcommand{\bk}{\boldsymbol{k}}
\newcommand{\bp}{\boldsymbol{p}}
\newcommand{\bq}{\boldsymbol{q}}
\newcommand{\balpha}{\ensuremath{\boldsymbol{\alpha}}}
\newcommand{\dd}[1]{\frac{\partial}{\partial #1}}
\newcommand{\Partder}[1]{\frac{\partial }{\partial #1}}
\newcommand{\partder}[2]{\frac{\partial #1}{\partial #2}}
\newcommand{\Cov}{\mathrm{Cov}}
\newcommand{\eff}{_{\mathrm{eff}}}
\newcommand{\SEp}{_{\mathrm{SEp}}}
\newcommand{\Irred}{\mathrm{Irred}}
\newcommand{\Sep}{\mathrm{Sep}}
\newcommand{\Nonsep}{\mathrm{Nonsep}}
\newcommand{\Irr}{_{\mathrm{Irr}}}
\newcommand{\RL}{_{\mathrm{L}}}
\newcommand{\QL}{_{\mathrm{QL}}}
\newcommand{\sgn}{\mathrm{sgn}}
\newcommand{\etc}{\mathrm{etc}}
\newcommand{\out}{\mathrm{out}}
\newcommand{\iin}{\mathrm{in}}
\newcommand{\inter}{\mathrm{int}}
\renewcommand{\rm}{\mathrm}
\newcommand{\mI}{\mathrm{I}}
\newcommand{\mS}{\mathrm{S}}
\newcommand{\mD}{\mathrm{D}}
\newcommand{\mG}{\mathrm{G}}
\newcommand{\mH}{\mathrm{H}}
\newcommand{\ext}{\mathrm{ext}}
\newcommand{\conn}{\mathrm{conn}}
\newcommand{\Counter}{\mathrm{Counter}}
\newcommand{\Linked}{\mathrm{linked}}
\newcommand{\linked}{\mathrm{linked}}
\newcommand{\mB}{\mathrm{B}}
\newcommand{\mC}{\mathrm{C}}
\newcommand{\SE}{\mathrm{SE}}
\newcommand{\G}{\mathrm{G}}
\newcommand{\sr}{\mathrm{sr}}
\newcommand{\op}{\mathrm{op}}
\newcommand{\opL}{\mathrm{opL}}
\newcommand{\cl}{\mathrm{cl}}
\newcommand{\clC}{\mathrm{clC}}
\newcommand{\bs}{\boldsymbol}
\newcommand{\bdot}{\boldsymbol\cdot}
\newcommand{\bnabla}{\bs{\nabla}}
\newcommand{\Q}{\mathcal{\boldsymbol{Q}}}
\newcommand{\bOm}{\mathcal{\boldsymbol{\Om}}}
\renewcommand{\H}{H}
\newcommand{\A}{A}
\newcommand{\bH}{\bs{\H}}
\newcommand{\bPsi}{\bs{\Psi}}
\newcommand{\bC}{\bs{C}}
\newcommand{\bR}{\bs{R}}
\newcommand{\bV}{\bs{V}}
\newcommand{\U}{U}
\newcommand{\Uop}{U_\op}
\newcommand{\Ucl}{U_\cl}
\newcommand{\p}{\hat{p}}
\newcommand{\hbp}{\hat{\bs{p}}}
\newcommand{\calH}{{\mathcal{H}}}
\newcommand{\calO}{\mathcal{O}}
\newcommand{\calG}{\mathcal{G}}
\newcommand{\calGop}{\mathcal{G}_\op}
\newcommand{\calGcl}{\mathcal{G}_\cl}
\newcommand{\calV}{\mathcal{V}}
\newcommand{\calL}{\mathcal{L}}
\newcommand{\dagg}{^{\dag}}
\newcommand{\intd}[1]{\int\frac{\dif #1}{2\pi}}
\newcommand{\half}{{\displaystyle\frac{1}{2}}}
\newcommand{\halfi}{{\displaystyle\frac{\im}{2}}}
\newcommand{\dint}{\int\!\!\!\int}
\newcommand{\ddint}{\dint\!\!\!\dint}
\newcommand{\dintd}[2]{\dint\frac{\dif #1}{2\pi}\,\frac{\dif #2}{2\pi}}
\newcommand{\ddintd}[4]{\ddint\frac{\dif #1}{2\pi}\,\frac{\dif #2}{2\pi}\,\frac{\dif #3}{2\pi}\,\frac{\dif #4}{2\pi}}
\newcommand{\ddt}{\frac{\partial}{\partial t}}
\newcommand{\intbr}{\int\dif\br\,}
\newcommand{\gamlim}{\ensuremath{\gamma\rightarrow 0}}
\newcommand{\vsp}{\vspace{0.5cm}}
\newcommand{\vvsp}{\vspace{0.25cm}}
\newcommand{\vvvsp}{\vspace{0.125cm}}
\newcommand{\mvsp}{\vspace{-0.5cm}}
\newcommand{\mvvsp}{\vspace{-0.25cm}}
\newcommand{\mvvvsp}{\vspace{-0.25cm}}
\newcommand{\mmvsp}{\vspace{-0.25cm}}
\newcommand{\mVsp}{\vspace{-1cm}}
\newcommand{\Vsp}{\vspace{1cm}}
\newcommand{\Wsp}{\vspace{2cm}}
\newcommand{\hsp}{\hspace{0.5cm}}
\newcommand{\hhsp}{\hspace{0.25cm}}
\newcommand{\Hsp}{\hspace{1cm}}
\newcommand{\HSP}{\hspace{2cm}}
\newcommand{\mmhsp}{\hspace{-0.25cm}}
\newcommand{\mhsp}{\hspace{-0.5cm}}
\newcommand{\mHsp}{\hspace{-1cm}}
\newcommand{\mHSP}{\hspace{-2cm}}
\newcommand{\ö}{\"{o}}
\newcommand{\Ö}{\"{O}}
\newcommand{\ä}{\"a}
\newcommand{\å}{\aa}
\newcommand{\Å}{\AA}
\newcommand{\hpsi}{\hat{\psi}}
\newcommand{\WU}{\widetilde{U}}
\renewcommand{\it}{\textit}
\newcommand{\bfit}[1]{\textbf{\it{#1}}}
\newcommand{\ul}{\underline}
\newcommand{\itul}[1]{\it{\ul{#1}}}
\newcommand{\abs}[1]{|{#1}|}
\newcommand{\Abs}[1]{\big|{#1}\big|}
\newcommand{\eq}{\eqref}
\newcommand{\rarr}{\rightarrow}
\newcommand{\larr}{\leftarrow}
\newcommand{\lrarr}{\leftrightarrow}
\newcommand{\LRarr}{\Longleftrightarrow}
\newcommand{\Rarr}{\Rightarrow}
\newcommand{\Lrarr}{\Longrightarrow}
\newcommand{\rr}{r_{12}}
\newcommand{\frr}{\frac{1}{r_{12}}}
\newcommand{\brr}{\br_{12}}
\newcommand{\DF}{D_{\rm{F}\nu\mu}}
\newcommand{\DFmn}{D_{\rm{F}\mu\nu}}
\newcommand{\SF}{S_{\rm{F}}}
\newcommand{\QED}{\rm{QED}}
\renewcommand{\sp}[2]{\bra{#1}{#2}\rangle}
\newcommand{\SP}[2]{\Bigbra{#1}{#2}\Big\rangle}
\newcommand{\qand}{\quad\rm{and}\quad}
\newcommand{\V}{V}
\newcommand{\VC}{V_\rm{C}}
\newcommand{\fC}{f_\rm{C}}
\newcommand{\VB}{V_\rm{B}}
\newcommand{\Vsf}{V_\rm{sp}}
\newcommand{\Usf}{U_\rm{sp}}
\newcommand{\Ssf}{S_\rm{sp}}
\newcommand{\Msf}{\calM_\rm{sp}}
\newcommand{\Ksf}{{\cal K_\rm{sp}}}
\newcommand{\MSE}{\calM_\rm{SE}}
\newcommand{\SSE}{S_\rm{SE}}
\newcommand{\USE}{U_\rm{SE}}
\newcommand{\calK}{\mathcal{\kappa}}
\newcommand{\calKc}{I_c}
\newcommand{\calF}{\mathcal{F}}
\newcommand{\calM}{{\mathcal{M}}}
\newcommand{\calR}{\mathcal{\hat{H}}}
\newcommand{\calB}{\mathcal{B}}
\newcommand{\GQ}{\Gamma_Q}
\newcommand{\bGQ}{\bs{\Gamma_Q}}
\newcommand{\Gk}{\bs{\Gamma_Q}}
\newcommand{\Gv}{\Gamma_Q}
\newcommand{\TD}{T_\rm{D}}
\newcommand{\F}{\rm{F}}
\renewcommand{\P}{\bs{P}}
\newcommand{\Util}{\widetilde{U}}
\newcommand{\Ucov}{U_\Cov}
\renewcommand{\S}{S}
\newcommand{\Vtil}{\widetilde{V}}
\newcommand{\Om}{\Omega}
\newcommand{\om}{\omega}
\newcommand{\OM}{\mathcal{\boldsymbol{\Om}}}
\newcommand{\Ombar}{\bar{\Omega}}
\newcommand{\VI}{V_{12}}
\newcommand{\VR}{V_{\rm{R}}}
\newcommand{\VRbar}{\bar{V}_{\rm{R}}}
\newcommand{\calE}{{\mathcal{E}}}
\newcommand{\CALE}{{\mathcal{E}}}
\newcommand{\la}{\lambda}
\newcommand{\FLL}{F_\rm{LL}}
\newcommand{\norm}[1]{||{#1}||}
\newcommand{\Norm}[1]{\big|\big|{#1}\big|\big|}
\newcommand{\NORM}[1]{\Big|\Big|{#1}\Big|\Big|}
\newcommand{\Fr}{Fr\'echet }
\newcommand{\Ga}{G\^ateaux }
\newcommand{\epsn}{\ensuremath{\epsilon}}
\newcommand{\Der}[1]{\frac{\dif}{\dif#1}}
\newcommand{\der}[2]{\frac{\dif#1}{\dif#2}}
\newcommand{\ave}[1]{\langle #1\rangle}
\newcommand{\Ave}[1]{\Big\langle #1\Big\rangle}
\newcommand{\bb}[2]{\big\{{#1}\,\big|\:{#2}\big\}}
\newcommand{\BB}[2]{\Big\{{#1}\,\Big|\:{#2}\Big\}}
\newcommand{\LL}{L^1\cap L^3}
\newcommand{\partdelta}[2]{\frac{\delta #1}{\delta #2}}
\newcommand{\pd}[1]{\frac{\delta #1}{\delta \calE}}
\newcommand{\Partdelta}[1]{\frac{\delta }{\delta #1}}
\newcommand{\partdeltan}[3]{\frac{\delta^#1 #2}{\delta #3^#1}}
\newcommand{\pdn}[2]{\frac{\delta^#1 #2}{\delta\calE^#1}}
 \newcommand{\ett}{^{(1)}}
\newcommand{\nol}{^{(0)}}
\newcommand{\PE}{P_\mathcal{E}}
\newcommand{\GE}{\Gamma_\E}
\newcommand{\GEP}{\Gamma_{\E'}}
\newcommand{\PEP}{P_{\calE'}}
\newcommand{\EP}{\calE'}
 \newcommand{\limgam}{\lim_{\gamlim}}
 \newcommand{\Ugam}[1]{U_\gamma(#1,-\infty)}
\newcommand{\Ugamtil}[1]{\widetilde{U}_\gamma(#1,-\infty)}
\newcommand{\Ugamt}{\widetilde{U}_\gamma}
\newcommand{\img}{\im\gamma}
\newcommand{\ime}{\im\eta}
\newcommand{\pbar}{\not{p}}
\newcommand{\Deltag}{\Delta_\gamma}

                                                  \newcommand{\tva}{^{(2)}}
                                                  \newcommand{\tre}{^{(3)}}
                                                  \newcommand{\fyr}{^{(4)}}
                                                  \newcommand{\enh}{^{(1/2)}}
                                                  \newcommand{\treh}{^{(3/2)}}
                                                  \newcommand{\femh}{^{(5/2)}}

\newcommand{\partdern}[3]{{\frac{\partial^#1 #2}
{\partial #3^#1}}}
\newcommand{\Partdern}[2]{{\frac{\partial^#1}
{\partial #2^#1}}}
  \newcommand{\MSC}{\rm{MSC}}
                                                  \newcommand{\Htil}{\widetilde{H}}
                                                  \newcommand{\Ubar}{\bar{U}}
                                                  \newcommand{\Hbar}{\bar{V}}
                                                  \newcommand{\Udot}{\dot{U}}
                                                  \newcommand{\Ubardot}{\dot{\Ubar}}
                                                  \newcommand{\Utildot}{\dot{\Util}}
                                                  \newcommand{\Cdot}{\dot{C}}
                                                  \newcommand{\Ombardot}{\dot{\Ombar}}
                                                  \newcommand{\bsdot}{\bs{\cdot}}

\newcommand{\npartdelta}[3]{\frac{\delta^#1 #2}{\delta #3^#1}}
\newcommand{\ip}[1]{| #1 \rangle \langle #1 |}
\newcommand{\con}{\mathrm{con}}
\newcommand{\ph}{\mathrm{ph}}
\newcommand{\bA}{\bs{A}}
\newcommand{\Pmu}{\partial^\mu}
\newcommand{\Pnu}{\partial^\nu}
\newcommand{\pmu}{\partial_\mu}
\newcommand{\pnu}{\partial_\nu}
\newcommand{\bAT}{\bs{A}_\perp}
\newcommand{\bAL}{\bs{A}_\parallel}
\newcommand{\bET}{\bs{E}_\perp}
\newcommand{\bEL}{\bs{E}_\parallel}
\newcommand{\bB}{\bs{B}}
\newcommand{\bE}{\bs{E}}
\newcommand{\bn}{\bs{n}}
\newcommand{\beps}{\bs{\eps}}
\newcommand{\HI}{H_\rm{int,I}}
\newcommand{\calHI}{{\cal H}_\rm{int,I}}
\newcommand{\ih}{\frac{\im}{\hbar}}
\newcommand{\bkx}{\bk\bdot\bx}
\newcommand{\halfS}{{\textstyle\frac{1}{2}\,}}
\newcommand{\h}{\hat{h}}
\newcommand{\DFnu}{D_{\rm{F}\nu\mu}}
\newcommand{\Eab}{E_{ab}}
\newcommand{\rE}{{\red E}}
                                                  \newcommand{\Ers}{E_{rs}}
                                                  \newcommand{\Etu}{E_{tu}}
                                                  \newcommand{\Eru}{E_{ru}}
                                                  \newcommand{\Ecd}{E_{cd}}
                                                  \newcommand{\Erd}{E_{rd}}
                                                  \newcommand{\epsi}{\epsilon}

\setlength{\unitlength}{0.75cm}

\title{The helium fine-structure controversy}
\author{Ingvar Lindgren\\ Physics Department, G\öteborg University,
G\öteborg, Sweden}
\date{\today}
\maketitle

\abstract There is presently disagreement between theory and
experiment as well as between different theoretical calculations
concerning the fine-structure splitting of the lowest $P$ state of
the neutral helium atom. We believe that we have found a minor
error in the formulas used by Drake et al. (Can. J. Phys. 80, 1195
(2002)) in their calculations, and we may have an explanation how
the error has occurred. To what extent this might resolve (part
of) the discrepancy is not known at present.

\section{Introduction}
The fine structure of the lowest $P$ state of neutral helium is of
great principal interest, since a comparison between theory and
experiment might yield an accurate and independent determination
of the fine-structure constant, $\alpha$. Unfortunately, various
theoretical calculations disagree, and there is also a significant
discrepancy between theory and experiment, when using an accurate
value of $\alpha$, determined in other ways, primarily from the
g-factor of the free electron. Even with $\alpha$ as a
free-running parameter, it is not possible to match theory and
experiment for the two fine-structure separations.

The most accurate experimental results are obtained by Gabrielse
\it{et al.}~\cite{Gabr05} and by Inguscio \it{et
al.}~\cite{Gius05}. The theoretical calculations have been
performed by Drake and coworkers~\cite{Drake02} as well as by
Pachucki and Sapirstein~\cite{PS00}.

The calculations of Drake \it{et al.} are based upon the works of
Sucher~\cite{Su58} and of Douglas and Kroll~\cite{DK74}. We have
now discovered that there is most likely a minor error in the
formulas of Douglas and Kroll and the corresponding formulas of
Zhang~\cite{Zhang96a} on which the works of Drake \it{et al.} are
based. This is estimated to lead to a correction of order
$\alpha^5$ Ry, which is beyond the accuracy of Douglas and Kroll
but might be relevant for the works of Drake and Zhang. We do not
know at present the magnitude of the effect and to what extent
this might resolve some of the discrepancies, but this is
certainly worth investigating. Below we shall review the analyses
of Douglas-Kroll (DK) and Zhang (Z), which are based on the early
work of Sucher (S)~\cite{Su58}, and point out where we believe the
mistake has been made.

\section{The analysis of Sucher, Douglas and Kroll}

In his thesis Sucher performed a perturbation expansion of the
Bethe-Salpeter equation (BSE)~\cite{SB51}, which with somewhat
different notations can be expressed
\begin{eqnarray}\label{BSA}
    &&\Psi(x,x')=\ddint\dif^4x_1\dif^4x_2\dif^4x'_1\dif^4x'_2\,\nn
    &&\times\;G'_0(x,x';x_2,x'_2)\,(-\im)\Sigma^*(x_2,x'_2;x_1,x'_1)\,\Psi(x_1,x'_1)
\end{eqnarray}
$G'_0$ is the zeroth-order two-particle Green's function, dressed
with all kinds of single-particle self-energy insertions, and
$\Sigma^*$ is the irreducible or proper two-particle self energy.
The function $G'_0$ is a product of two single-particle Green's
functions, satisfying the relation
\begin{equation}\label{Gdef1}
    \Big(\im\Partder{t}-h_1\Big)G(x,x_0)=\im\delta^4(x-x_0)
\end{equation}
which leads to (S 1.6, DK 2.19)
\begin{eqnarray}\label{BSB}
    \Big(\im\Partder{t}-h_1\Big) \Big(\im\Partder{t'}-h_2\Big)\Psi(x,x')
    =\im\dint\dif^4x_1\dif^4x'_1\, \Sigma^*(x,x';x_1,x'_1)\,\Psi(x_1,x'_1)
\end{eqnarray}
where $h_{1,2}$ are the Dirac single-particle Hamiltonians.

With the wave function being of the form
\begin{equation}\label{Timedep}
    \Psi(x,x')=\Psi(T,\tau,\bx,\bx')=\me^{-\im ET}\,\Psi(\tau,\bx,\bx')
\end{equation}
where $T=(t+t')/2$ is the average time and $\tau=t-t'$ is the
relative time, the BSE can after a Fourier transform be expressed
in operator form (S 1.12, DK 2.26, Z 1)
\begin{equation}  \label{BSD}
  \calF\,\Psi(\epsilon)=g\,\Psi(\epsilon)
\end{equation}
Here,
\begin{equation}  \label{calF}
  \calF=\Big(E/2+\epsilon-h_1\Big) \Big(E/2-\epsilon-h_2\Big)
\end{equation}
and
\begin{equation}  \label{gPsi}
  g\,\Psi(\epsilon)=\frac{\im}{2\pi}\,
  \Sigma^*(\epsilon)\,\Phi
\end{equation}
where  (S 1.31, DK 3.7, Z 14)
\begin{equation}\label{Phi}
    \Phi=\int\dif{\epsilon}\,\Psi(\epsilon)
\end{equation}
is the "\it{equal-time}" function.

The interaction $g$ can be separated into a Coulomb part and a
"remainder"
\begin{equation}
  g=g_c+g_\Delta
\end{equation}
leading to
\begin{equation}  \label{BSE}
  \Psi(\epsilon)=\big(\calF-g_\Delta\big)^{-1}g_c\,\Psi(\epsilon)
\end{equation}
From \eqqref{gPsi} it follows that (S 1.32), (DK 3.8)
\begin{equation}\label{Ic}
  g_c\Psi(\epsilon)=\frac{\im}{2\pi}\,I_c\,\Phi
\end{equation}
where $I_c$ is the Coulomb interaction.

Integrating \eqqref{Ic} over the relative energy $\epsilon$, leads
to (S 1.34, DK 3.10)
\begin{equation}  \label{BSF}
  \Phi=\im\intd{\epsilon}\big(\calF-g_\Delta\big)^{-1}I_c\,\Phi
\end{equation}

The inverse of the operator $\calF$ is essentially a product of
two electron propagators, yielding
\begin{equation}
    \im\intd\epsilon\,\calF^{-1}=-G_0(E)=
    \frac{1}{E-h_1-h_2}\,\big(\Lambda_{++}-\Lambda_{--}\big)
\end{equation}
where $\Lambda_{++},\,\Lambda_{--}$ are two-particle projection
operators for doubly positive and negative states, respectively.
This leads to the equation (S 1.47), (DK 3.26)
\begin{equation}\label{BS3}
  \boxed{\Big[h_1+h_2+\big(\Lambda_{++}-\Lambda_{--}\big)\calKc+
  \im\intd{\epsilon}\,D\calF^{-1}g_\Delta({\cal
  F}-g_\Delta)^{-1}\calKc\Big]\Phi=E\,\Phi}
\end{equation}
where
\begin{equation}\label{D}
  D=E-h_1-h_2
\end{equation}
This is the starting point for the further analysis.

\section{Perturbation expansion of the BS equation}
The operator on the left-hand side of \eqqref{BS3} can be separated
into a \it{no-pair Coulomb operator}
\begin{subequations}
\begin{equation}\label{Dc}
    H_c=h_1+h_2+\Lambda_{++}\calKc\Lambda_{++}
\end{equation}
a \it{Coulomb virtual-pair operator}
\begin{equation}  \label{QED1}
  H_{\Delta1}=\Lambda_{++}\calKc(1-\Lambda_{++})-\Lambda_{--}\calKc
  \end{equation}
\it{and a relativity and transverse photon operator}
\begin{equation}  \label{QED2}
  H_{\Delta2}=\im\intd{\epsilon}D\calF^{-1}g_\Delta({\cal
  F}-g_\Delta)^{-1}\calKc
\end{equation}
\end{subequations}

Starting from the no-pair approximation
\begin{equation}  \label{DCeq}
  H_c\,\Psi_c=E_c\,\Psi_c
\end{equation}
the Brillouin-Wigner expansion yields the energy contribution
\begin{equation}
  \label{BW1}
  \Delta E=E-E_c=\bra{\Psi_c}V+V\Gamma V+V\Gamma V\Gamma V+\cdots\ket{\Psi_c}
  \end{equation}
where
\begin{equation}
  \label{BWRes}
  \Gamma=\GQ(E)=\frac{Q}{E-H_c}
\end{equation}
is the resolvent. This leads to the expansion terms (S 2.19-21, DK
3.43, Z 28)
 \begin{subequations}
\begin{equation}
  \Delta E^{(1)}=\bra{\Psi_c}H_\Delta\ket{\Psi_c}
\end{equation}
\begin{equation}
  \Delta E^{(2)}=\bra{\Psi_c}H_\Delta\Gamma H_\Delta\ket{\Psi_c}
\end{equation}
\begin{equation}
  \Delta E^{(3)}=\bra{\Psi_c}H_\Delta\Gamma H_\Delta\Gamma H_\Delta\ket{\Psi_c}
\end{equation}
\end{subequations}
etc.

It can be shown that there is no first-order contribution from
$H_{\Delta1}$, and the first-order energy contribution then
becomes (DK 3.44)
\begin{equation}  \label{E1}
  \Delta E^{(1)}=\bra{\Psi_c}H_{\Delta2}\ket{\Psi_c}=
  \bra{\Psi_c}\im\intd{\epsilon}D\calF^{-1}J\calF^{-1}\calKc\ket{\Psi_c}
\end{equation}
where
\begin{equation}  \label{J}
  J=g_\Delta(1-\calF^{-1} g_\Delta)^{-1}
\end{equation}

Of special interest here is one of the second-order contributions
(DK 3.46, note some misprints)
\begin{equation}  \label{E2b}
  \Delta E_b^{(2)}=\bra{\Psi_c}H_{\Delta1}\,\Gamma\,H_{\Delta2}\ket{\Psi_c}=
  \bra{\Psi_c}\calKc\Lambda_{--}\,\Gamma\,\im\intd{\epsilon}D\calF^{-1}J\calF^{-1}
  \calKc\ket{\Psi_c}
\end{equation}
It can easily be shown that $\Lambda_{--}\,\Gamma D=\Lambda_{--}$
(DK 3.41). Using the relation \eq{Dc} and with $D_c=E-c-h_1-h_2$,
we have $E_c-H_c=D_c-\Lambda_{++}D_c\Lambda_{++}$, and the no-pair
equation \eq{DCeq} can be written (DK 3.51)
\begin{equation}  \label{DCeq2}
  (D_c-\Lambda_{++}\calKc)\,\Psi_c=0
\end{equation}
Then the second-order correction $\Delta E_b^{(2)}$ \eq{E2b} can
be expressed
\begin{equation}  \label{E2b2}
  \Delta E_b^{(2)}=\bra{\Psi_c}(\calKc-D_c)\,
  \im\intd{\epsilon}\calF^{-1}J\calF^{-1}\calKc\ket{\Psi_c}
  \end{equation}
This can be combined with the first-order correction $\Delta
E^{(1)}$ \eq{E1}, yielding
\begin{equation}  \label{E1E2b}
  \bra{\Psi_c}(D+\calKc-D_c)\,\im\intd{\epsilon}\calF^{-1}J\calF^{-1}\calKc
  \ket{\Psi_c}=\bra{\Psi_c}(\calKc+\Delta E)\,\im\intd{\epsilon}\calF^{-1}J\calF^{-1}\calKc
  \ket{\Psi_c}
\end{equation}
Here, the $\Delta E$ term differs in sign from (DK 3.54) and (Z
37).

The reason for the discrepancy between our result here and those
of Douglas and Kroll and of Zhang seems to be that the latter use
the relation
\begin{equation}  \label{G0mod}
  \calF^{-1}=S_1S_2\equiv \big(S_1+S_2\big)\big(S_1^{-1}+S_2^{-1}\big)^{-1}
  =D^{-1}\big(S_1+S_2\big)
\end{equation}
where $S_{1,2}$ are electron propagators, and the identity (DK
3.50a, note misprint)
\begin{equation}\label{Dm1}
  D^{-1}=\frac{1}{D_c}-\frac{\Delta E}{D_cD}
\end{equation}
to transform the first-order equation \eqref{E1} to
\begin{eqnarray}  \label{E1a}
   \mhsp\Delta
   E^{(1)}=\bra{\Psi_c}D_c^{-1}(1-\Delta E/D)\,\im\intd{\epsilon}(S_1+S_2)
   J(S_1+S_2)I_c\ket{\Psi_c}
\end{eqnarray}
and the second-order correction \eqref{E2b2} to
\begin{equation}  \label{E1E2b}
   \Delta E_b^{(2)}=\bra{\Psi_c}(\calKc-D_c)\,D^{-2}\,\im\intd{\epsilon}(S_1+S_2)
   J(S_1+S_2)I_c\ket{\Psi_c}
\end{equation}
Then they cancel $D_c^{-1}$ in the first equation against
$D_c\,D^{-2}$ in the second. Zhang approximates $D$ by $D_c$ in
all $\Delta E\ett$ and $\Delta E\tva$ expressions, which leads to
the same cancellation. According to \eqqref{Dm1}, this leads to an
error of $2\Delta E/(D_cD)$, which explains the difference.

The difference is of the form $2\Delta E\times\Delta E\ett$, which
is of order $\alpha^5$ Ry. The correction $\Delta E$ represents
the difference between the full energy and the Dirac-Coulomb
energy, which contains the instantaneous Breit interaction and
therefore of order $\alpha^2$ Ry. This interaction does not
contribute to the fine-structure splitting, and therefore $\Delta
E\ett$ is order $\alpha^3$ Ry, making the correction of order
$\alpha^5$ Ry. This does not affect the work of Douglas and Kroll,
who study corrections up to $\alpha^4$ Ry, while it is of
relevance for Zhang and Drake who go one step further.

\section{Equal-time approximation}
Another possible source of the discrepancy between theory and
experiment might be the fact that the calculations -- following
Sucher -- are based upon the so-called \it{equal-time
approximation}, where the particles are assumed to have the same
time. This is in contrast to the covariant Bethe-Salpeter
equation, where the particles have individual times. It is hard to
tell what size an effect of this kind might have on the results
and to what extent it can be visible at the present level of
accuracy. Most likely, however, this will be the case at some
level.

\section*{Acknowledgements}
The author wishes thank his coworkers Sten Salomonson and Daniel
Hedendahl as well as his international colleagues Gordon Drake and
Gerald Gabrielse for stimulating discussions on this subject.

\bibliographystyle{c:/Sty/prsty}
\bibliographystyle{unsrt}

\end{document}